\documentclass[twocolumn, a4paper,fleqn,usenatbib]{aastex63}

\received{\today}
\revised{}
\accepted{}

\usepackage{newtxtext,newtxmath}
\usepackage[T1]{fontenc}
\usepackage{ae,aecompl}

\usepackage{graphicx}	% Including figure files
\usepackage{amsmath}	% Advanced maths commands
\usepackage{amssymb}	% Extra maths symbols
\usepackage{bm}
\usepackage{times}

\newcommand{\cf}{cf.,~}
\newcommand{\ie}{i.e.,~}
\newcommand{\eg}{e.g.,~}

\usepackage{todonotes}

\shorttitle{GW170817 and GW190814: tension on the maximum mass}
\shortauthors{Nathanail, Most and Rezzolla}
\begin{document}

\title{GW170817 and GW190814: tension on the maximum mass}

\author{Antonios Nathanail}
\affiliation{Institut f\"ur Theoretische Physik, Max-von-Laue-Strasse
  1, 60438 Frankfurt, Germany}
\author[0000-0002-0491-1210]{Elias R. Most}
\affiliation{Princeton Center for Theoretical Science, Princeton University, Princeton, NJ 08544, USA}
\affiliation{Princeton Gravity Initiative, Princeton University, Princeton, NJ 08544, USA}
\affiliation{School of Natural Sciences, Institute for Advanced Study, Princeton, NJ 08540, USA}
\author[0000-0002-1330-7103]{Luciano Rezzolla}
\affiliation{Institut f\"ur Theoretische Physik, Max-von-Laue-Strasse
  1, 60438 Frankfurt, Germany}
\affiliation{Frankfurt Institute for Advanced Studies, Ruth-Moufang-Strasse 1,
  60438 Frankfurt, Germany}
\affiliation{School of Mathematics, Trinity College, Dublin 2, Ireland}

% These dates will be filled out by the publisher
%\date{Accepted XXX. Received YYY; in original form ZZZ}

% Enter the current year, for the copyright statements etc.
%\pubyear{2017}

% Don't change these lines

%%\begin{document}
%\label{firstpage}
%%\pagerange{\pageref{firstpage}--\pageref{lastpage}}
%\maketitle

\begin{abstract}
The detection of the binary events GW170817 and GW190814 has provided
invaluable constraints on the maximum mass of nonrotating configurations
of neutron stars, $M_{_{\rm TOV}}$. However, the large differences in the
neutron-star masses measured in GW170817 and GW190814 has also lead to a
significant tension between the predictions for such maximum masses, with
GW170817 suggesting that $M_{_{\rm TOV}} \lesssim 2.3\,M_{\odot}$, and
GW190814 requiring $M_{_{\rm TOV}} \gtrsim 2.5\,M_{\odot}$ if the
secondary was a (non- or slowly rotating) neutron star at merger. Using a
genetic algorithm, we sample the multidimensional space of parameters
spanned by gravitational-wave and astronomical observations associated
with GW170817. Consistent with previous estimates, we find that all of
the physical quantities are in agreement with the observations if the
maximum mass is in the range $M_{_{\rm TOV}} = 2.210^{+0.116}_{-0.123}
\,M_{\odot}$ within a $2\textrm{-}\sigma$ confidence level. By contrast,
maximum masses with $M_{_{\rm TOV}} \gtrsim 2.5\,M_{\odot}$, not only
require efficiencies in the gravitational-wave emission that are well
above the numerical-relativity estimates, but they also lead to a
significant under-production of the ejected mass. Hence, the tension can
be released by assuming that the secondary in GW190814 was a black hole
at merger, although it could have been a rotating neutron star before.
\end{abstract}


%%%%%%%%%%%%%%%%%%%%%%%%%%%%%%%%%%%%%%%%%%%%%%%%%%

%%%%%%%%%%%%%%%%% BODY OF PAPER %%%%%%%%%%%%%%%%%%

%-----------------------------------------------------------------------
\section{Introduction}
\label{sec:intro}
%-----------------------------------------------------------------------

The recent detection of the gravitational-wave (GW) event GW190814 has
seen involved the merger of black hole (BH), with a mass of $22.2-24.3\,
M_{\odot}$, with a compact object having a much smaller mass of
$2.50-2.67\,M_{\odot}$ \citep{Abbott2020b}. The unclear nature of the
secondary component has raised questions about the astrophysical
evolutionary paths that would yield objects with these masses in a binary
system. When assigning a NS nature to the secondary in GW190814, two
scenarios are possible. In the first one, the secondary was a nonrotating
or slowly rotating NS at merger, so that GW190814 should effectively be
considered a BH-NS merger \citep[see, \eg][for some possible
    formation scenarios]{Zevin2020, Safarzadeh2020,
    Kinugawa2020,Liu2020,Lu2020,Ertl2020}. In
this case, the maximum mass $M_{_{\rm TOV}}$ of nonrotating NSs needs to
reach values as large as $\gtrsim 2.5\,M_{\odot}$ \citep{Fattoyev2020,
  Sedrakian2020, Tan2020, Tsokaros2020, Godzieba2020, Biswas2020}. In the
second scenario, the need for a large maximum mass can be replaced by the
presence of rapid rotation. In fact, it has been shown that uniformly
rotating NS can support about $20\%$ more mass than nonspinning ones
\citep{Breu2016, Shao2020}. Note that in the case of NSs with a phase
transition, universal relations are still present, but depend on the
properties of the phase transition \citep{Bozzola2019, Demircik2020}.

Based on these universal relations, \citet{Most2020d} and
\citet{Zhang2020} have pointed out that a massive (rapidly) rotating NS
with a mass $>2.5\, M_\odot$ is perfectly consistent with a maximum mass
$M_{_{\rm TOV}} \simeq 2.3\,M_{\odot}$ inferred from the GW170817 event
\citep[see, \eg][]{Rezzolla2017, Shibata2019}. Given the difficulty of
sustaining rapid rotation over the very long timescales associated with
the inspiral of the binary, the secondary must have collapsed at one
point before merger, so that in this second scenario GW190814 should
effectively be considered a BH-BH merger.

While a priori both scenarios are plausible, shedding light on which of
them is the most likely is important from several points of view. To this
scope, we here exploit the rich variety of GW and electromagnetic
observables that have been obtained with GW170817 to explore the two
scenarios combining the constraints set from the GW and electromagnetic
signal from GW170817. In particular, we employ a genetic algorithm to
sample through the distributions of maximum masses, ejected matter
\citep{Drout2017, Cowperthwaite2017, Kasen2017, Villar2017,
  Coughlin2018}, and GW emission from numerical-relativity (NR)
simulations \citep{Zappa2018}. Consistent with previous results
\citep{Rezzolla2017,Shibata2019} we find that GW170817's observations
clearly set an upper limit for the maximum mass of $M_{_{\rm TOV}}
\lesssim 2.33\,M_{\odot}$. When forcing the algorithm to allow for
maximum masses $M_{_{\rm TOV}} \gtrsim 2.4\, M_\odot$, we find that this
requires unrealistically large GW efficiencies from the merger remnant
and a deficit in the ejected matter.

%-----------------------------------------------------------------------
\section{Framework for the genetic algorithm}
\label{sec:setup}
%-----------------------------------------------------------------------

The observations of a bright blue kilonova has provided convincing
evidence that the merger remnant in GW170817 could not have collapsed
promptly to a BH. Rather, it must have survived for a timescale of the
order of one second \citep{Gill2019, Lazzati2020, Hamidani2020}, and
sufficiently large so that the hypermassive NS (HMNS) produced by the
merger has reached uniform rotation at least in its core
\citep{Margalit2017,Rezzolla2017}. Following \citet{Rezzolla2017}, we
recall that quasi-universal relations exist between the masses of
uniformly rotating stellar models along the stability line to BH
formation and the corresponding dimensionless angular momentum $j_{\rm
  coll}$ normalised to the maximum (Keplerian) one $j_{\rm Kep}$
\citep{Breu2016}. We here express this relation as
\begin{align}
\hspace{-0.8cm}
  \chi(j_{\rm coll}/j_{\rm Kep}) := \frac{M_{\rm crit}}{M_{_{\rm TOV}}}
  = 1 + \alpha_2 \left( \frac{j_{\rm coll}}{j_{\rm Kep}} \right)^2
  + \alpha_4 \left( \frac{j_{\rm coll}}{j_{\rm Kep}} \right)^4 \, ,
\label{eq:jcol}
\end{align}
where $\alpha_2 = 1.316\times10^{-1}$ and $\alpha_4= 7.111\times10^{-2}$,
and the value of the Keplerian specific angular momentum is approximately
given by $j_{\rm Kep} \sim 0.68$ \citep[see Eq. (4) in][for more accurate
  estimates]{Most2020d}. The function $\chi$ is defined between 0 and 1
and describes all models with a mass that is critical for collapse to a
BH. To fix ideas, in the case of nonorotating models, $j_{\rm coll}=0$
and $\chi(0)=1$, while for maximally rotating models $j_{\rm coll}=j_{\rm
  Kep}$ and \citep{Breu2016} $\chi(1) := {M_{\rm max}}/{M_{_{\rm TOV}}}
\approx 1.20_{-0.05}^{+0.02}$, where $M_{\rm max}$ is the maximum mass
that can be sustained through uniform rotation \citep[see][for
  differentially rotating stars]{Weih2017}. Note that range $\chi(1)$ is
based on a specific set of hadronic equations of state (EOSs) and that a
different estimate suggests $\chi(1)=1.17_{-0.05}^{+0.02}$
\citep{Shao2020}.

Because Eq. \eqref{eq:jcol} expresses a relation between gravitational
masses, while the electromagnetic emission from GW170817 informs us about
the ejected baryonic mass, we need a relation between gravitational and
baryonic mass $M_b$ for uniformly rotating NSs at the mass-shedding limit
\citep[see, \eg][for a detailed discussion]{Timmes1996,
    Gao2020}. Also in this case, this relation obeys a quasi-universal
relation near the values of the maximum mass that, with a
$2\textrm{-}\sigma$ uncertainty, is given by $\eta := {M_{b, {\rm
      max}}}/{M_{\rm max}}\approx 1.171\pm0.014$ at the maximum-mass
limit \citep{Rezzolla2017}. Note that $\eta$ is in principle a function
of $M$ and that the value reported above is for $M=M_{\rm max}$. However,
$\eta$ is almost constant in the neighbourhood of $M_{\rm max}$ -- where
all of our considerations are made -- so that hereafter we simply write
the conversion between the two masses as $M_b = \eta M$.

The total gravitational mass of GW170817 as inferred from the GW signal
is $M_g =2.73_{-0.01}^{+0.04}$ \citep{Abbott2018a}, whose corresponding
baryonic mass $M_b$ \textit{soon} after the merger can be thought of as
being given by the combination of the baryonic mass in the HMNS $M_{b,
  {_{\rm HMNS}}}$ -- itself composed of the mass in the core and in a
Keplerian disk -- and of the mass ejected dynamically, \ie
\begin{equation}
  M_b =  M_{b, {\rm core}} + M_{b, {\rm disk}} + M_{\rm ej}^{\rm dyn} =  \eta M^{*}_g\,,
\end{equation}
where $M^*_g := M_g - M_{\rm GW}^{\rm insp}$ and $M_{\rm GW}^{\rm insp}$
is the energy lost to GWs in the inspiral. Here, the last equality
relates the baryonic and gravitational mass of the merger
remnant. Defining now $\xi$ as the fraction of the HMNS baryonic mass in
the core, the two components of the HMNS \textit{shortly after} merger
can be written as
\begin{equation}
  M_{b, {\rm core}} := \xi \left(M_b - M_{\rm ej}^{\rm dyn}\right) =
  \xi \left(\eta M^{*}_g - M_{\rm ej}^{\rm dyn}\right)
 \label{eq:core}
\end{equation}
The fraction $\xi$ is in principle unknown, but numerical simulations
have shown that this ratio is actually weakly dependent on the EOS and
given by $\xi \approx 0.95^{+0.04}_{-0.05}$ \citep{Hanauske2016}. As time
goes by, the merger remnant will loose part of its baryonic mass via the
emission of magnetically driven or viscous-driven winds, so that at
collapse it will have a baryonic mass
\begin{equation}
\hspace{-0.8cm}
  M_{b} = M_{b, {\rm core}}^{\rm coll} + M_{b, {\rm
  disk}}^{\rm coll} + M_{\rm ej}^{\rm dyn} + M_{\rm ej}^{\rm blue} + M_{\rm
    ej}^{\rm red} \,,
\end{equation}
where the last equality follows from rest mass conservation and $M_{b,
  {\rm core}}^{\rm coll}$ and $M_{b, {\rm disk}}^{\rm coll}$ are the
respective values of the core and the disk at the time when BH formation
of the core is triggered, while $M_{\rm ej}^{\rm blue}$ ($M_{\rm ej}^{\rm
  red}$) is the part of the ejected matter leading to the blue (red)
emission in the kilonova and differs from the dynamical ejecta from the
timescale over which the material is lost. The two components also differ
in the typical velocities of the matter, which is larger in the blue
component ($v/c \lesssim 0.3$ for the blue part and $v/c \ll 0.1$ for the
red part), but also with in the electron fraction $Y_e$, which is again
larger in the blue component ($0.2 \lesssim Y_e \lesssim 0.3$ for the
blue part and $Y_e \lesssim 0.2$ for the red part). Numerical
simulations of remnant disks indicate that most of the red ejecta
originate from the disk, whereas most of the blue ejecta will come from
the hot surface of the HMNS. Hence, for simplicity we will assume that
the blue ejecta originate from the HMNS only, while the red ejecta
exclusively represent unbound material of the disk. We classify the
latter via a parameter
\begin{align}
   f_{\rm disk} := M_{\rm ej}^{\rm red} / M_{b, {\rm disk}} \simeq 0.2 -
   0.5 \,,
  \label{eq:red}
\end{align}
representing the unbound fraction of the disk mass, which can be
estimated based on numerical simulations \citep{Siegel2017,
  Fernandez2018, Fujibayashi2018, Nathanail2020b}. In a merger scenario
as that of GW170817, where the remnant may have lived for about one
second \citep{Gill2019}, BH formation is triggered when the gravitational
mass is reduced by the emission of GWs and the remnant hits the stability
line for uniformly rotating models with a massive core $M_{b, \rm
  core}^{\rm coll}$
\begin{align}
\hspace{-0.8cm}
  M_{b, \rm core}^{\rm coll} &= \eta M^{*}_g - M_{b,{\rm disk}}
  - M_{\rm ej}^{\rm dyn} - M_{\rm ej}^{\rm blue} = \eta \chi M_{_{\rm
      TOV}} \,,
  \label{eq:mbcoll}
\end{align}
where the last equality relates the baryon mass of the remnant core to
the maximum mass $M_{_{\rm TOV}}$ of nonrotating NS via
Eq. \eqref{eq:jcol}. Indeed, when expressed as as a constraint equation
on the maximum mass, Eq. \eqref{eq:mbcoll} can also be written
\begin{align}
  \eta \chi M_{_{\rm TOV}}  = \xi \left( \eta M^{*}_{g} - M_{\rm ej}^{\rm
  dyn} \right) - M_{\rm ej}^{\rm blue} \,.
  \label{eq:mtov}
\end{align}

Two more equations can be used for consistency
\begin{align}
  \label{eq:constraints_a}
\hspace{-0.8cm}
  M_{\rm ej}^{\rm red} &= f_{\rm disk} \left( 1-\xi \right) 
  \left( \eta M^{*}_{g} - M_{\rm ej}^{\rm
  dyn} \right)\,, \\
  \label{eq:constraints_b}
\hspace{-0.8cm}
  \chi M_{_{\rm TOV}} & = M^{*}_g - \eta^{-1}\!\left(
  M_{b,{\rm disk}}^{\rm coll} \!+\!  M_{\rm ej}^{\rm dyn} \!+\! M_{\rm ej}^{\rm blue} \!+\!
  M_{\rm ej}^{\rm red} \right)
  - M^{\rm post}_{_{\rm GW}}\,,
\end{align}
where the first one expresses the conservation of rest-mass leading to
the kilonova emission and the second one the conservation of
gravitational mass since $M^{\rm post}_{_{\rm GW}}$ is the mass lost to
GWs after the merger. Expression \eqref{eq:constraints_b} does not
constrain $M^{\rm post}_{_{\rm GW}}$, which thus remains
undetermined. As a way around (see also \citet{Fan2020}), we use
an approximate quasi-universal relation between the total mass lost to
GWs $M_{\rm GW}^{\rm tot}$ and the specific angular momentum of the
remnant after the merger \citep{Zappa2018}
\begin{align}
  m_{\rm GW}^{\rm tot} \sim c_0 + c_1 j_{\rm rem, 20} +
  c_2 \left(j_{\rm rem, 20}\right)^2 \,,
  \label{eq:egwtot}
\end{align}
where $m_{\rm GW}^{\rm tot} := M_{\rm GW}^{\rm tot}/({M_g \nu})$, $j_{\rm
  rem, 20} := J_{\rm rem, 20}/(M_g^2\nu)$ is the specific angular
momentum of the remnant within $\sim 20\,{\rm ms}$ from the merger, and
$\nu := m_1 m_2/(m_1+m_2)^2$ is the symmetric mass ratio. Note that
$c_0=0.9, c_1=-0.4$ and $c_2=0.05$ \citep{Zappa2018} and that the two
component masses $m_1$ and $m_2$ are chosen considering the low-spin
prior for GW170817, \ie $\nu \in [0.243,0.25]$. By splitting total mass
lost to GWs into the two components relative to the inspiral and
post-merger, \ie $M_{\rm GW}^{\rm tot} = M^{\rm insp}_{_{\rm GW}} +
M^{\rm post}_{_{\rm GW}}$, Eq. \eqref{eq:egwtot} allows us to introduce
an additional constraint between $j_{\rm coll}$ -- which we derive from
Eq. \eqref{eq:jcol} -- and $M^{\rm post}_{_{\rm GW}}$.

Two more remarks before concluding the presentation of our
methodology. First, not all of the merger remnant's angular momentum will
end up in the collapsed object. A number of physical processes will move
part of the angular momentum outwards, placing it on stable orbits
relative to the newly formed BH. Because the efficiency of this process
depends on microphysics that is poorly understood, we account for this by
introducing a fudge factor $f_B$ defined as $j_{\rm coll} =: (1 - f_B)
j_{\rm rem, 20}$, so that the specific angular momentum of the disk is
$j_{\rm disk} := f_B j_{\rm rem, 20}$. Second, since in
Eqs. \eqref{eq:mtov}, \eqref{eq:constraints_a} and
\eqref{eq:constraints_b} the function $\chi$ always appears together with
$M_{_{\rm TOV}}$, it is difficult to set reasonable ranges for
$\chi$. However, numerical simulations have revealed that the
dimensionless spin of the BH produced by the merger $j_{_{\rm BH}}$ (and
hence $j_{\rm coll} \lesssim j_{_{\rm BH}}$) is actually constrained in a
rather limited range, \ie $0.7 \lesssim j_{_{\rm BH}} \lesssim 0.9$
\citep{Kastaun2013, Bernuzzi2013}. Exploiting this information, and
assuming conservatively that $80\%$ of the specific angular momentum at
collapse is inherited by the BH, \ie $j_{\rm coll}=0.8\,j_{_{\rm BH}}$,
we can effectively constrain $\chi$ to be in the range $1.11 \leq \chi
\lesssim 1.22$. Very similar results are obtained when making the more
drastic assumption that only half of the BH spin comes from the remnant,
\ie $j_{\rm coll}=0.5\,j_{_{\rm BH}}$, which further reduces the lower
limit to be $\chi=1.05$ (see Supplement Material for details).

In summary, we need to solve a multidimensional parametric problem as
expressed by Eqs. \eqref{eq:mtov}, \eqref{eq:constraints_a} and
\eqref{eq:constraints_b} after varying in the appropriate ranges the
(ten) free parameters in the system: $\chi, \xi, \eta, M_{_{\rm GW}}^{\rm
  insp}, M_{\rm ej}^{\rm blue}, M_{\rm ej}^{\rm dyn}, M_{_{\rm TOV}},
\nu, f_{\rm disk}$ and $f_B$. While we treat all these parameters
equally, some of them ($\eta, M_{\rm ej}^{\rm dyn}, \nu$), vary in very
narrow ranges and their variations do not significantly affect the
overall results. In practice, at any iteration of the genetic algorithm
we ensure that:
\textit{(i)} the total gravitational mass of the system is $M_g
=2.73_{-0.01}^{+0.04}$ \citep{Abbott2017,Abbott2018a};
\textit{(ii)} the total ejected mass is $M_{\rm ej}^{\rm tot}= 0.0537
\,\pm \,0.013\, M_{\odot}$ \citep{Arcavi2017, Nicholl2017, Chornock2017,
  Cowperthwaite2017, Villar2017, Drout2017, Kasen2017, Tanaka2017,
  Waxman2017,Coughlin2018};
\textit{(iii)} the dynamically ejected mass is $M_{\rm ej}^{\rm
  dyn}\approx 10^{-3}\,M_{\odot}$ \citep{Sekiguchi2015, Bovard2017,
  Radice2018a, Poudel2020};
\textit{(iv)} the blue/red ejected components are respectively $0.005 <
M_{\rm ej}^{\rm blue}/M_{\odot} < 0.02$ and $0.03 < M_{\rm ej}^{\rm
  red}/M_{\odot} <0.55$ \citep{Drout2017, Cowperthwaite2017, Smartt2017,
  Kasliwal2017, Kasen2017, Villar2017, Tanaka2017, Waxman2017,
  Coughlin2018} we have also explored a larger upper bound on the
  blue ejecta, \ie $M_{\rm ej}^{\rm blue}/M_{\odot} < 0.05$, finding no
  significant difference; see Supplemental Material;
\textit{(v)} the maximum mass is taken to be in the range
$1.70\,M_{\odot}<M_{_{\rm TOV}}<3\,M_{\odot}$, note that the posterior 
lower bound is consistent with pulsar observations \citep{Antoniadis2013,
Cromartie2019}; 
\textit{(vi)} the energy radiated in GWs before the merger is constrained
to be $0.035 \lesssim M_{_{\rm GW}}^{\rm insp}/M_{\odot} \lesssim 0.045$
\citep{Zappa2018}. Note that all of the priors discussed in points
\textit{(i)}--\textit{(vi)} are uniform.

%-----------------------------------------------------------------------
\section{Results}
\label{sec:results}
%-----------------------------------------------------------------------

\begin{figure}
  \centering
  \includegraphics[width=0.45\textwidth]{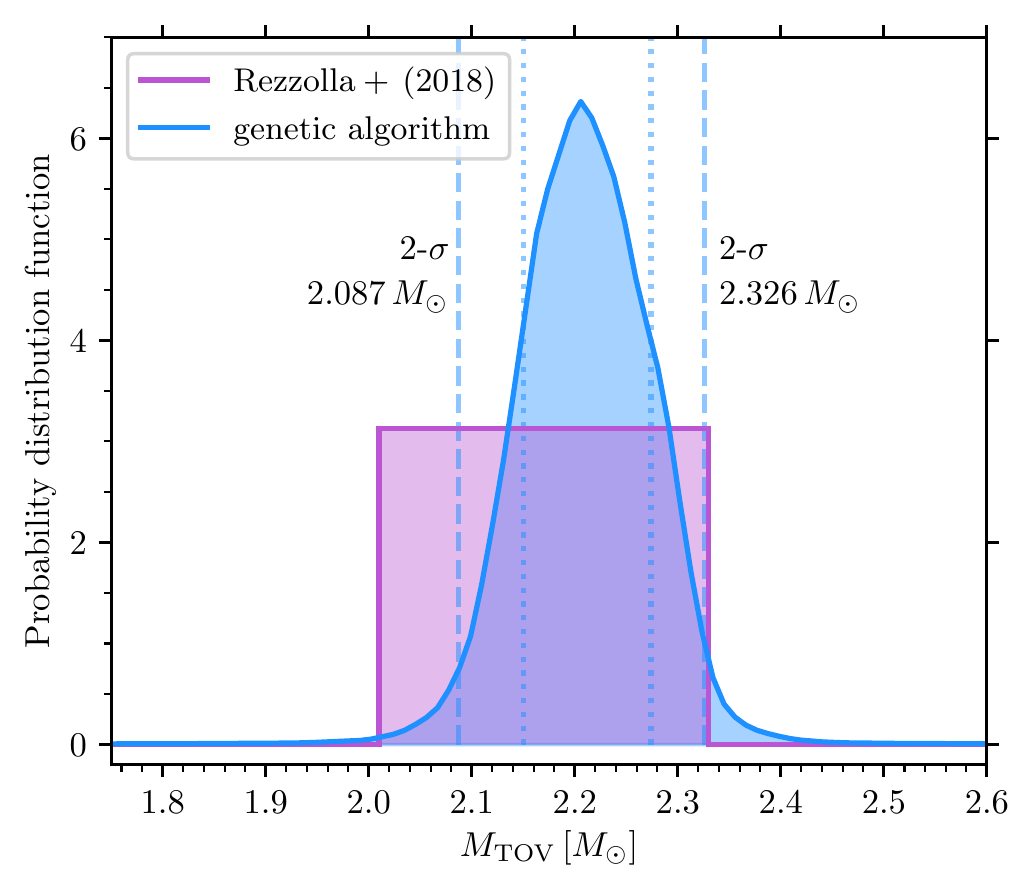}
        \caption{Uniform posterior from the analysis of
          \citet{Rezzolla2017} (magenta) and the posterior obtained with
          the multidimensional genetic algorithm (blue) discussed
          here. Indicated with vertical lines are the $1\textrm{-}\sigma$
          (dotted) and $2\textrm{-}\sigma$ (dashed) values.}
  \label{fig:mtov}
\end{figure}

\begin{figure*}
  \centering \includegraphics[width=0.45\textwidth]{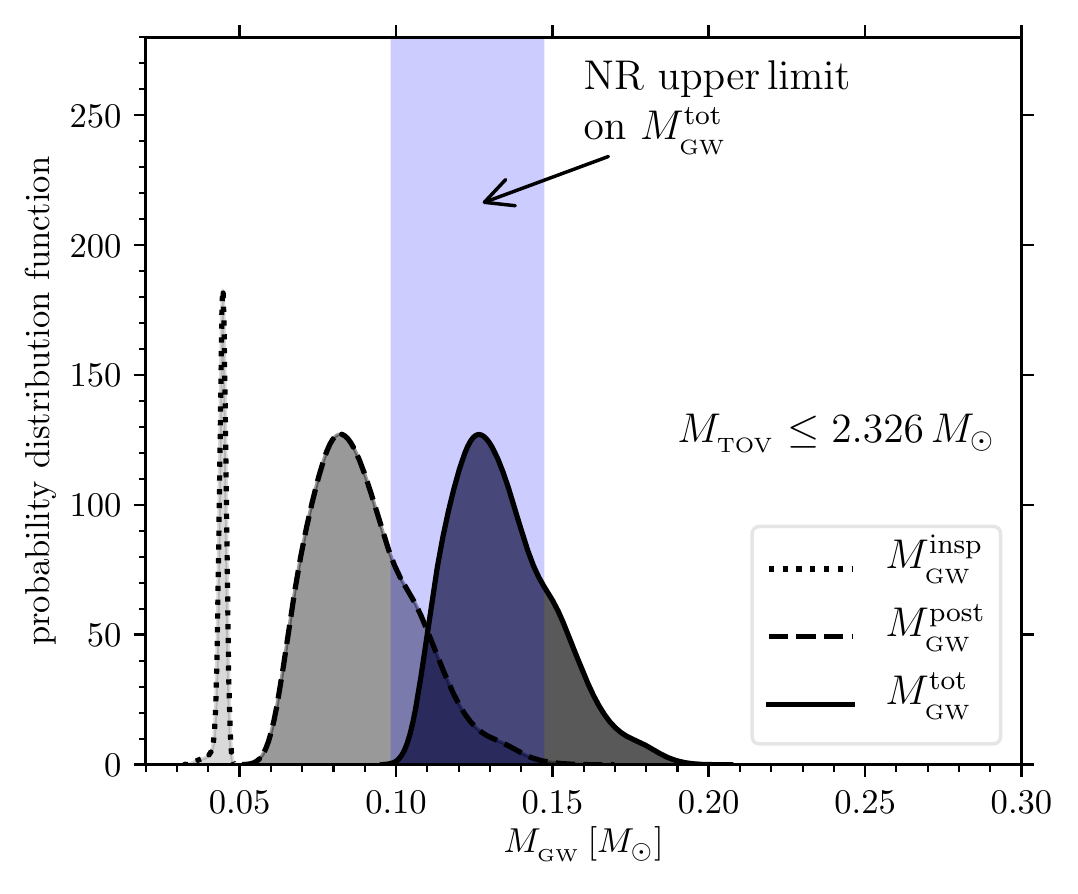}
  \hspace{0.75cm}
  \centering \includegraphics[width=0.45\textwidth]{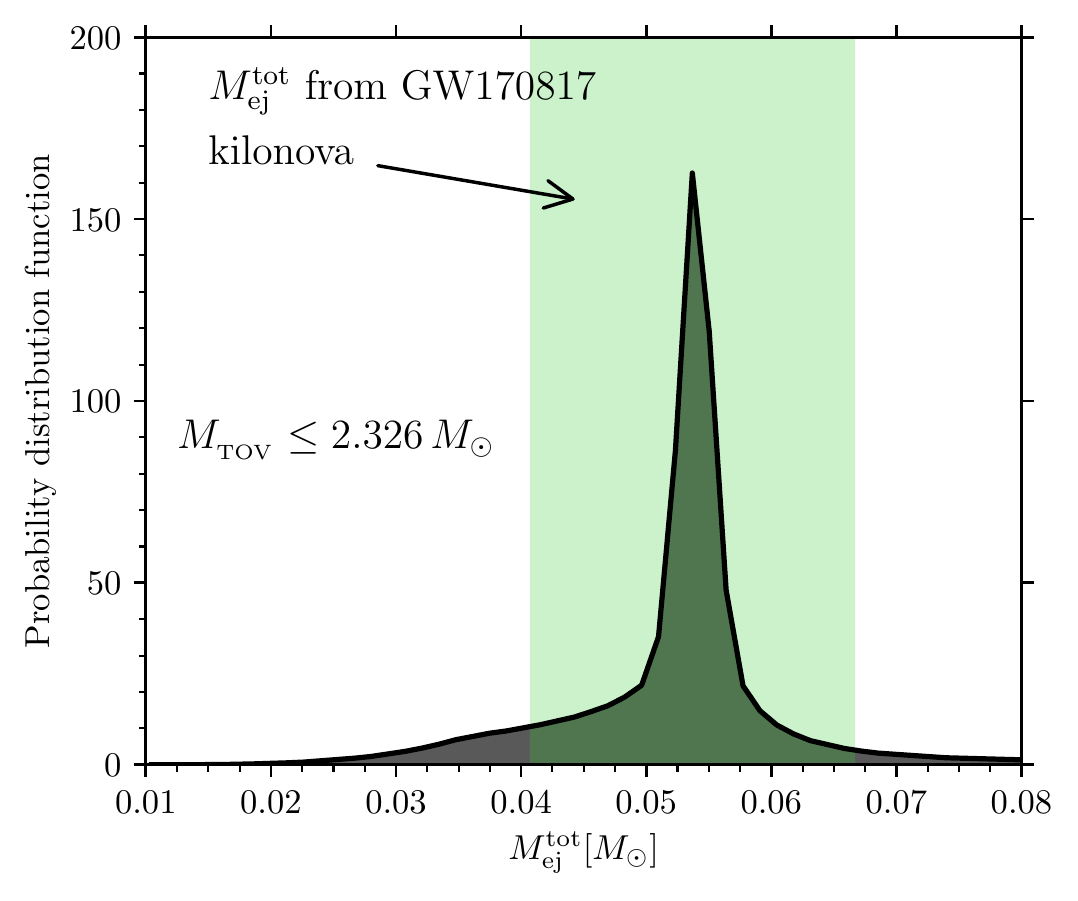}
  \caption{\textit{Left panel:} Posteriors for the mass radiated in GWs
    consistent with the distribution in Fig. \ref{fig:mtov}; the
    lavender-shaded area reports the upper limit coming from NR
    simulations. \textit{Right panel:} Posterior for the total ejected
    mass consistent with the distribution in Fig. \ref{fig:mtov}; the
    green-shaded area reports the range estimated in the literature.}
  \label{fig:ej}
\end{figure*}

Figure \ref{fig:mtov} provides a first important impression of the
results of the genetic-algorithm. In particular, shown with a
magenta-shaded area is the maximum-mass estimate made by
\cite{Rezzolla2017} and which is a simple uniform posterior for $M_{_{\rm
    TOV}}=2.16_{-0.16}^{+0.17}\,M_{\odot}$. Shown instead with a
blue-shaded area is the posterior distribution obtained with the genetic
algorithm. The median of the distribution is $M_{_{\rm TOV}}=
2.210^{+0.117}_{-0.123}\,M_{\odot}$, where the errors reported here are
for $2\textrm{-}\sigma$ uncertainty. Overall, this yields a lower bound
of $M_{_{\rm TOV}}>2.087\,M_{\odot}$ and an upper bound of $M_{_{\rm
    TOV}}<2.326\,M_{\odot}$ at $2\textrm{-}\sigma$ level (vertical dashed
lines), and thus in good agreement with massive-pulsar measurements
\citep{Antoniadis2013, Cromartie2019} and previous estimates
\citep{Rezzolla2017,Shibata2019}. Interestingly, our results are
  in good agreement with the conclusions reached by \citet{Shao2020} and
  \citet{Fan2020}, who have also considered the post-merger GW emission
  to deduce bounds on the maximum mass.

As a consistency check, we can use set of parameters that yield the
maximum-mass distribution in Fig. \ref{fig:mtov}, to estimate the GW
energy lost both in the inspiral and in the post-merger. This is shown in
the left panel of Fig. \ref{fig:ej}, where we report the posterior
distributions for $M_{_{\rm GW}}^{\rm insp}$ (black dotted line) and
$M_{\rm GW}^{\rm post}$ (black dashed line), as well as their sum,
$M_{_{\rm GW}}^{\rm tot}$ (black solid line). Also marked with a vertical
lavender-shaded is the upper limit estimated by \cite{Zappa2018}, $E^{\rm
  tot}_{\rm GW} /M_g \leq 0.045\,M_{\odot} c^2$, on the basis of a large
number of NR simulations, with an associated uncertainty of $20\%$. A
similar consistency is found in the right panel of Fig. \ref{fig:ej},
where we report the posterior of the total ejected mass consistent with
the maximum-mass distribution in Fig. \ref{fig:mtov}. Indicated with a
green-shaded area are the constraints obtained from the kilonova
observations of GW170817. More specifically, the width of the shaded area
represents the standard deviation using various estimates for the total
ejected mass $M_{\rm ej}^{\rm tot}$ estimated for GW170817
\citep{Arcavi2017, Nicholl2017, Chornock2017, Cowperthwaite2017,
  Villar2017, Drout2017, Kasen2017, Tanaka2017, Waxman2017,
  Coughlin2018}. Clearly, also the ejected-mass distribution is in
perfect agreement with observational bounds when the maximum mass is
below $2.326\,M_{\odot}$.

\begin{figure*}
  \centering
  \includegraphics[width=0.45\textwidth]{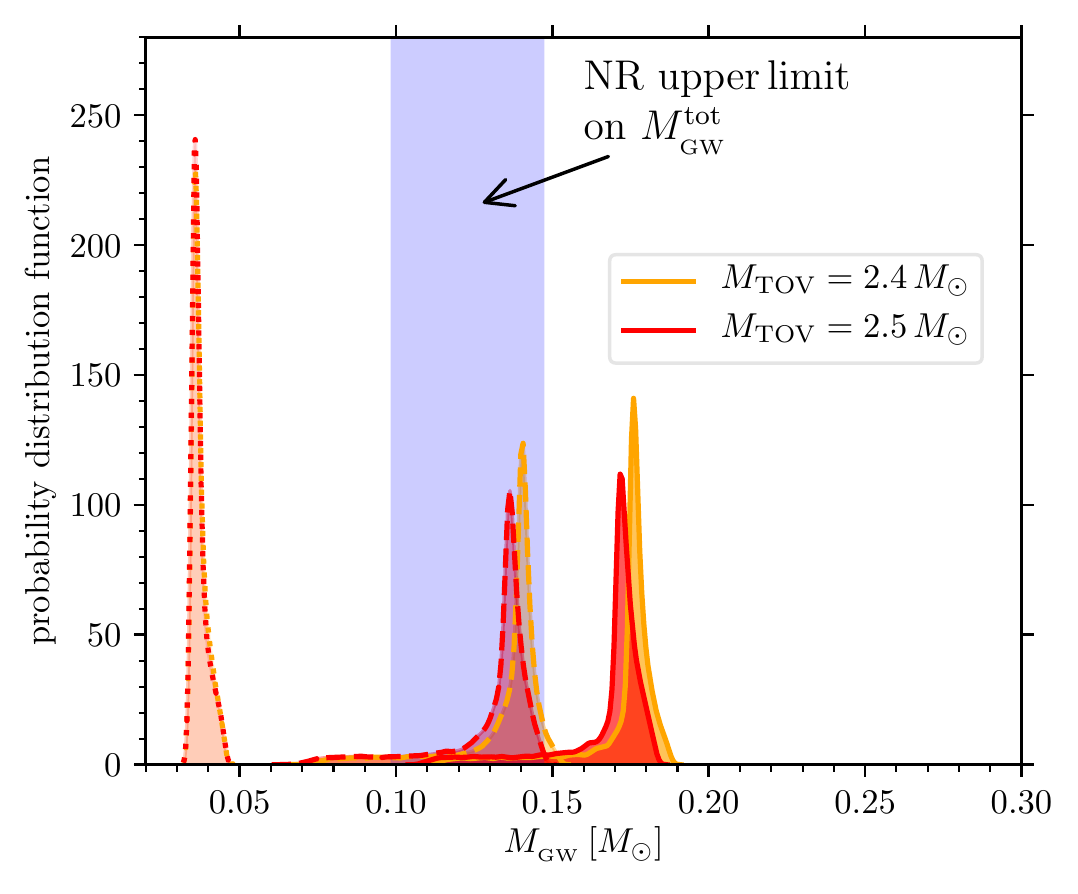}
  \hspace{0.75cm}
  \includegraphics[width=0.45\textwidth]{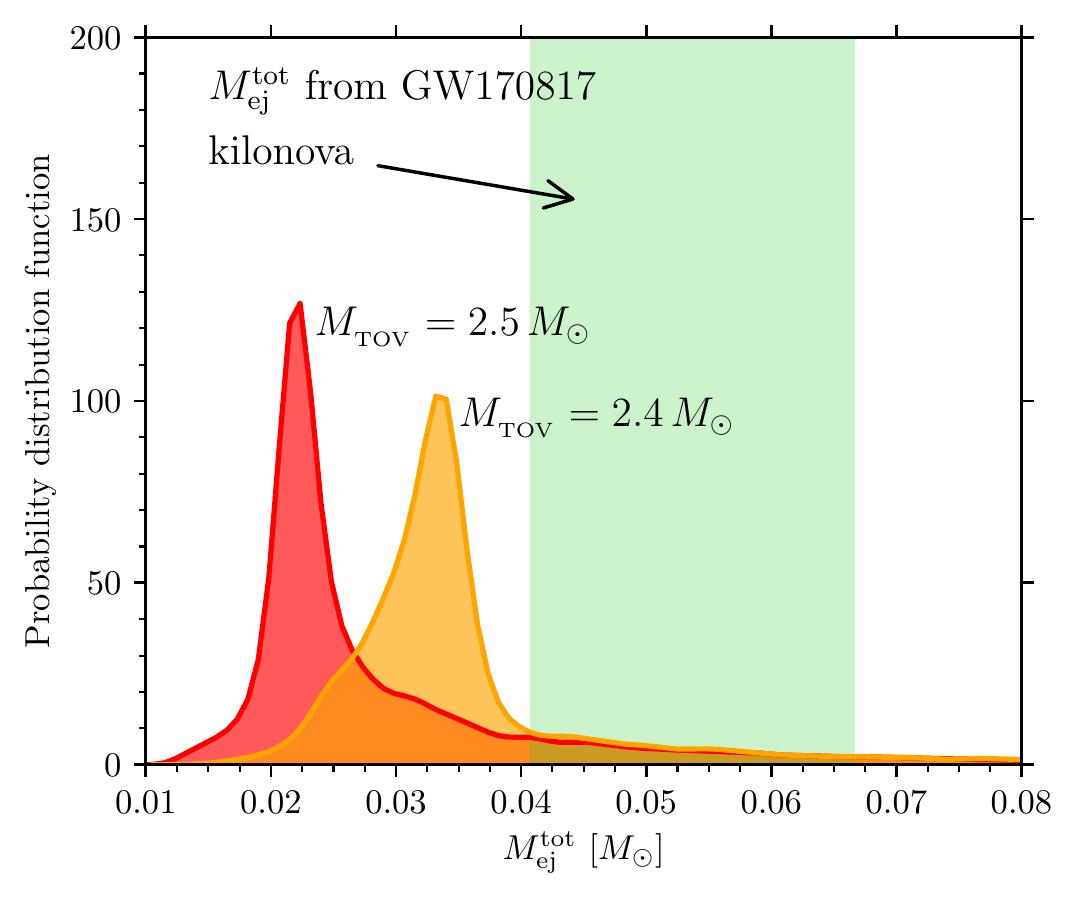}
  \caption{\textit{Left panel:} The same as in the left panel of
    Fig. \ref{fig:ej}, but when considering two fixed values for the
    maximum mass, \ie $M_{_{\rm TOV}} = 2.4\,M_{\odot}$ (orange) and
    $M_{_{\rm TOV}} = 2.5\,M_{\odot}$ (red). \textit{Right panel:} The
    same as in the right panel of Fig. \ref{fig:ej}, but for two fixed
    maximum-mass values. }
  \label{fig:breakdown}
\end{figure*}

Given these results, it is natural to ask: \textit{does anything break
  down when larger maximum masses are considered?} The positive answer to
this question is contained in Fig. \ref{fig:breakdown}, whose panels are
similar to those in Fig. \ref{fig:ej}, when however the genetic algorithm
is forced to consider two specific values of the maximum mass, namely,
$M_{_{\rm TOV}} = 2.4\,M_{\odot}$ and $M_{_{\rm TOV}} =
2.5\,M_{\odot}$. Concentrating first on the left panel of
Fig. \ref{fig:breakdown}, it is clear that when allowing for large
maximum masses, the mass radiated in GWs after the merger, $M_{_{\rm
    GW}}^{\rm post}$ (red dashed line), is significantly larger than what
NR simulations predict; this is true both for $M_{_{\rm TOV}} =
2.4\,M_{\odot}$ and for $M_{_{\rm TOV}} = 2.5\,M_{\odot}$. This behaviour
is due to the fact that remnants with a given $\chi$ will radiate more
GWs if they are more massive [\cf Eq. \eqref{eq:constraints_b}]. Next,
when considering the right panel of Fig. \ref{fig:breakdown} it is also
easy to realize that large maximum masses lead to a deficit in the
ejected matter. This is simply due to the fact that considering
large-mass stars inevitably reduces the portion of the budget available
for the ejecta. We have confirmed that, even if (unrealistically) large
additional angular momentum transport was assumed, these results remain
unchanged for $M_{_{\rm TOV}} \gtrsim 2.5\, M_\odot$, (see Supplemental
Material).

In summary, while a value $M_{_{\rm TOV}} \lesssim 2.3\,M_{\odot}$ is
fully consistent with the GW emission from NR simulations and the
observed ejected mass, a value $M_{_{\rm TOV}} \gtrsim 2.5\,M_{\odot}$
requires efficiencies in the GW emission that are well above the
estimates from a large number of accurate NR simulations and, overall,
leads to an under-production of ejected mass.

%-----------------------------------------------------------------------
\section{Conclusions}
\label{sec:conclusions}
%-----------------------------------------------------------------------

We have carried out a systematic investigation to ascertain whether the
tension on the maximum mass following the detections of GW170817 and
GW190814 can in some way be resolved or at least attenuated. In
particular, we have employed a genetic algorithm to sample through the
multidimensional space of parameters that can be built on the basis of
the astronomical observations (\ie ejected mass in the various
components), GW observations (\ie gravitational masses of the binary
components), and of NR simulations (\ie properties of the remnant and
efficiency of GW emission).

The results of this investigation have allowed us to refine in a
probabilistic manner the previous estimates of the maximum mass
\citep{Rezzolla2017}, obtaining that $M_{_{\rm TOV}}=
2.210^{+0.117}_{-0.123}\,M_{\odot}$ within a $2\textrm{-}\sigma$
confidence level. In this range, all of the physical quantities are in
very good agreement with the estimates coming from the observations. By
contrast, we find that considering maximum masses with $M_{_{\rm TOV}}
\gtrsim 2.4-2.5\,M_{\odot}$ requires efficiencies in the GW emission well
above the NR estimates and leads to a significant under-production of the
ejected mass, well below the values expected from the
observations. Although robust, our results can be strengthened in a
number of ways. Improved post-merger modeling and long-term NR
simulations \citep{Fujibayashi2018} can help to narrow the uncertainties
in the parameter ranges for $\chi$ and $j_{\rm rem}$. Refined universal
relations of uniformly rotating NSs including temperature dependence
\citep{Koliogiannis2020b}, will also help narrowing down the errors in
$\eta$ and the upper limit of $\chi$. Such improvements will be crucial
to understand the viability of the maximum-mass constraint for $M_{\rm
  TOV} \lesssim 2.3\,M_\odot$.

In light of these considerations, we conclude that the secondary in
GW190814 was most likely a BH at merger, although it may well have been a
rotating NS at some stage during the evolution of the binary system.

\smallskip\noindent\textit{Acknowledgements.~}It is a pleasure to thank
C. Ecker, J. Papenfort, and L. Weih for useful comments. Support comes in
part also from ``PHAROS'', COST Action CA16214 and the LOEWE-Program in
HIC for FAIR. ERM gratefully acknowledges support from a joint fellowship
at the Princeton Center for Theoretical Science, the Princeton Gravity
Initiative and the Institute for Advanced Study.

\software{\texttt{Scipy} \citep{SciPy2020},
          \texttt{Corner} \citep{Foreman2016},
          }

%%%%%%%%%%%%%%%%%%%%%%%%%%%%%%%%%%%%%%%%%%%%%%%%%%



\bibliographystyle{yahapj}
%\bibliography{aeireferences}

%%%%%%%%%%%%%%%%%%%%%%%%%%%%%%%%%%%%%%%%%%%%%%%%%%

%%%%%%%%%%%%%%%%% APPENDICES %%%%%%%%%%%%%%%%%%%%%

\newpage
\section*{Supplemental Material}

In what follows we provide additional information that complements the
one provided in the main text. While the details illustrated below do not
vary the conclusions drawn in the main text, they provide additional
technical details on the genetic algorithm employed in our analysis. In
addition, they help investigate how the results change when the
parameters are varied beyond the (reasonable) ranges assumed so far.

For the solution of our multidimensional parametric problem the procedure
we adopt is as follows \citep[see also][for additional
  information]{Fromm2019, Nathanail2020b}. We start by recalling that
genetic algorithms are designed to generate high-quality solutions to
problems of this type where a searching optimization is sought. The name
follows from the operators of mutation, crossover and selection that are
normally found in biological systems. Our choice of a genetic algorithm
in place of a more traditional Bayesian analysis based on a Markov-Chain
Monte Carlo approach is motivated mostly by the overall simplicity of our
problem and the reduced computational costs that are associated with a
genetic algorithm.

In practice, our algorithm samples through the parameter space of the ten
free parameters. From those it computes the $M_{\rm ej}^{\rm red}$
through Eqs. \eqref{eq:core} and \eqref{eq:red}, $M^{\rm post}_{_{\rm
    GW}}$ through Eq. \eqref{eq:egwtot} and the specific angular momentum
at collapse $j_{\rm coll}$ via Eq. \eqref{eq:jcol}, using the sampled
value of$\chi$. Subsequently, Eqs. \eqref{eq:mtov},
\eqref{eq:constraints_a}, and \eqref{eq:constraints_b} are solved to
match the observed values of $M_g$ and $M_{\rm ej}^{\rm tot}$ within the
errors, finding the best-fit values. The genetic algorithm employed here
makes use of Python packages from the \texttt{SciPY} software library
\citep{SciPy2020}.

\begin{figure*}
  \centering
  \includegraphics[width=0.95\textwidth]{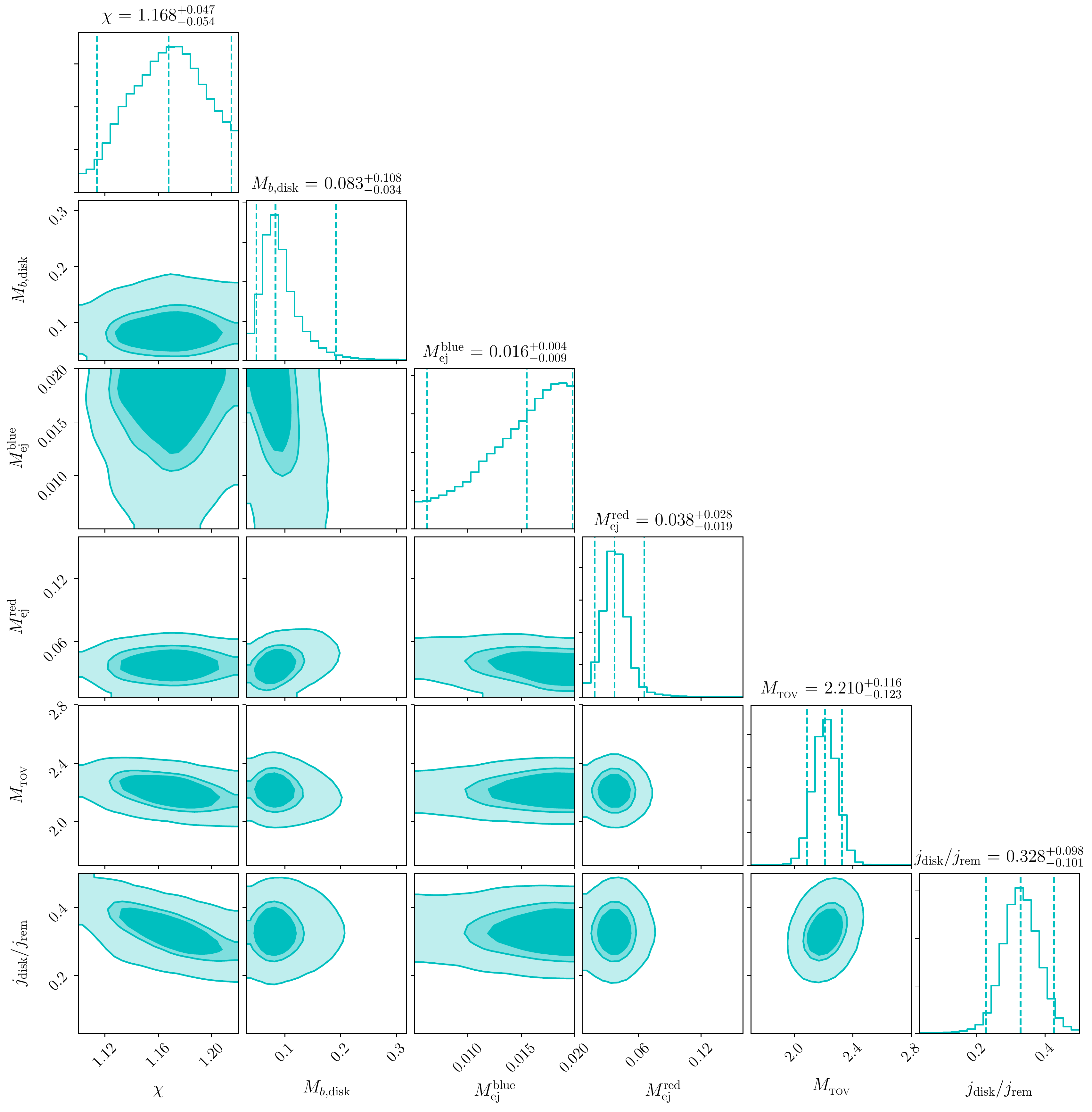}
	\caption{Corner plot reporting the posterior distributions of the
          most important parameters in our analysis. Indicated with the
          two outermost vertical dashed lines are the corresponding
          $2\textrm{-}\sigma$ values, while the labels on the diagonal
          cells report the average values (central vertical dashed
          line).}
  \label{fig:ej_cp}
\end{figure*}

\begin{figure*}
  \centering
  \includegraphics[width=0.95\textwidth]{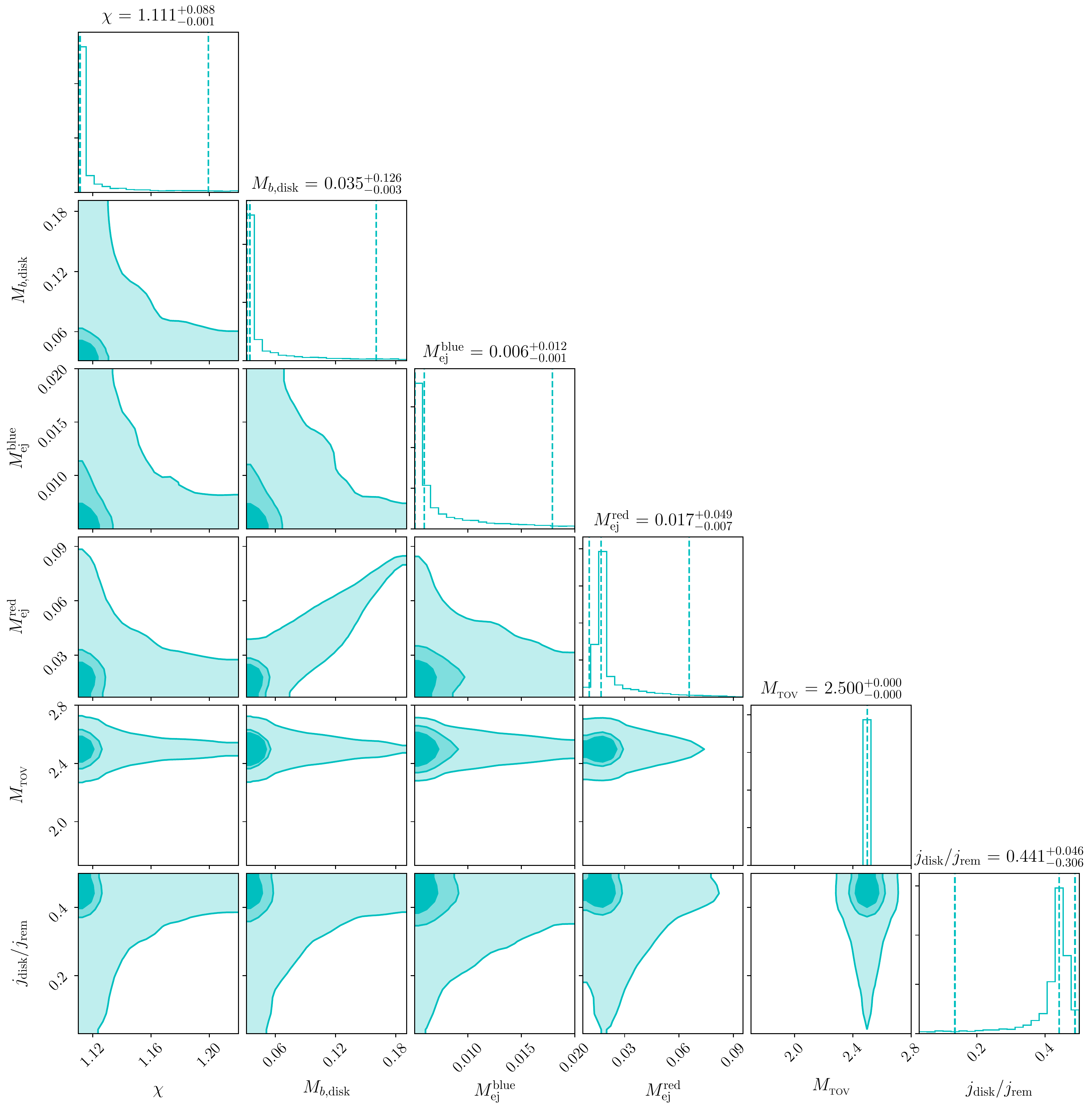}
  \caption{Same as Fig. \ref{fig:ej_cp} but when the maximum mass
          is held fixed at the value $M_{_{\rm TOV}} = 2.5\,M_{\odot}$.}
  \label{fig:ej_cp2}
\end{figure*}

As a corollary to the discussion made in the main text and relative to
Figs. \ref{fig:mtov}--\ref{fig:breakdown}, we provide with the corner
plots in Fig. \ref{fig:ej_cp} information on the probability distribution
functions of the various quantities involved in our analysis. More
specifically, Fig. \ref{fig:ej_cp} shows the corner plot relative to
maximum-mass posterior shown in Fig. \ref{fig:mtov} and should therefore
accompany the information presented in Fig.  \ref{fig:ej}. On the other
hand, Fig.  \ref{fig:ej_cp2} refers to the case when the genetic
algorithm is forced to consider $M_{_{\rm TOV}} = 2.5\,M_{\odot}$. In
this case, the maximum-mass is set to vary uniformly in the very small
interval around $M_{_{\rm TOV}} = 2.5^{+0.0001}_{-0.0001}\,M_{\odot}$,
leaving all the other parameters free to be adjusted till a best-fit
is found. In this sense, the information in Fig. \ref{fig:ej_cp2}
complements what is reported in Fig. \ref{fig:breakdown} and shows that all
the posterior distributions are pushed to be very narrow at the edges of
the allowed ranges. For instance, the dimensionless spin $\chi$ is
narrowly peaked around its minimum value $1.1$, the mass in the disk is
much smaller and of the order of $\simeq 0.035\,M_{\odot}$, while the
blue and red ejecta are comparable and equal to $\simeq
0.019\,M_{\odot}$.

Note that to avoid having a large number of small panels, we have limited
ourselves either to the most salient ones, omitting those quantities for
which the distributions are either almost constant or restricted to a
very small region. More specifically, in Fig. \ref{fig:ej_cp} the values
found are: $\xi = 0.973_{-0.035}^{+0.011}$, $\eta =
1.171_{-0.013}^{+0.014}$, $M_{\rm ej}^{\rm dyn} =
0.001_{-0.001}^{+0.001}\,M_{\odot}$, $f_{\rm disk} =
0.437_{-0.209}^{+0.061}$, and $\nu = 0.243_{-0.000}^{+0.005}$, which
corresponds to $q = 0.719_{-0.006}^{+0.136}$. We have also
  explored a modified scenario in which the blue ejecta are larger than
  inferred from observations. In particular, we have adjusted the upper
  bound on the blue ejecta from $M_{\rm ej}^{\rm blue}/M_{\odot} < 0.02$
  to $M_{\rm ej}^{\rm blue}/M_{\odot} < 0.05$. In this case, we find that
  the blue ejecta converge to a distribution with a median around $\sim
  0.03\, M_{\odot}$, while the red ejecta component decreases to $\sim
  0.02\, M_{\odot}$. At the same time, the changes in the posterior for
  the maximum mass are minute, \ie $M_{_{\rm TOV}} =
  2.192^{+0.125}_{-0.092} \,M_{\odot}$.

\begin{figure*}
  \centering
  \includegraphics[width=0.45\textwidth]{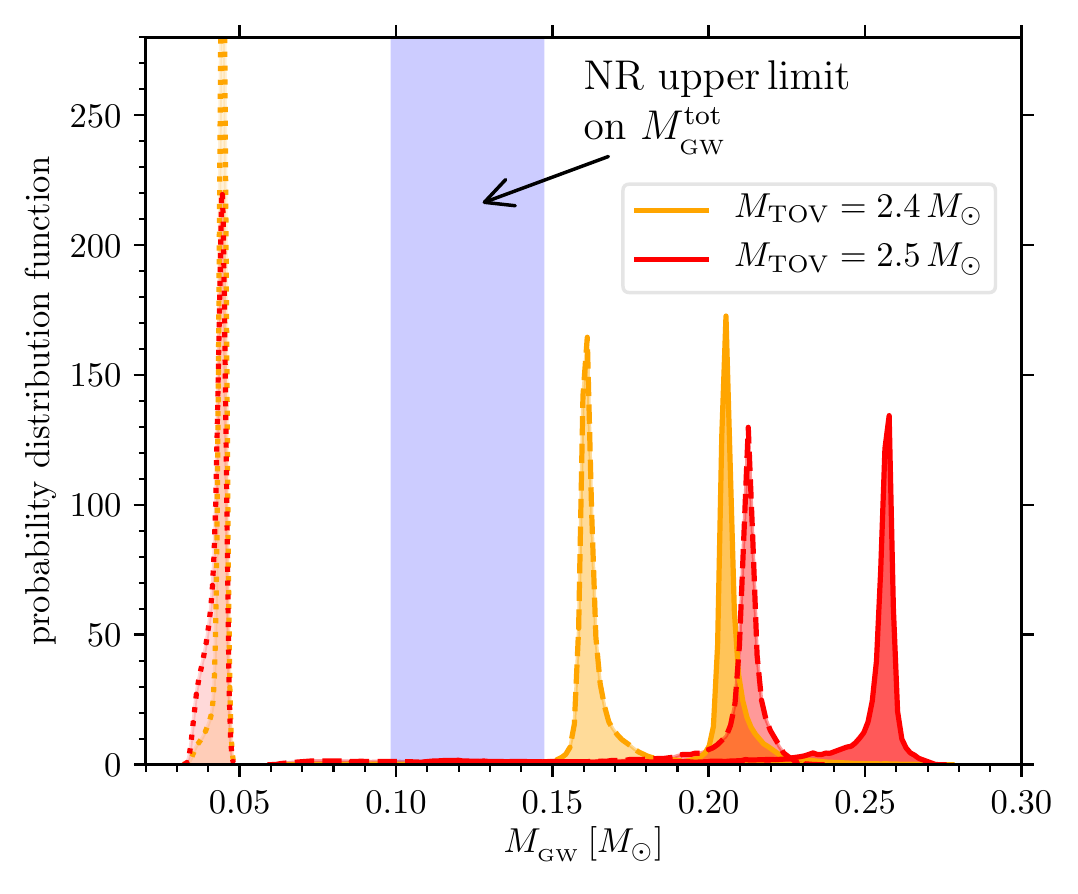}
  \hspace{0.75cm}
  \includegraphics[width=0.45\textwidth]{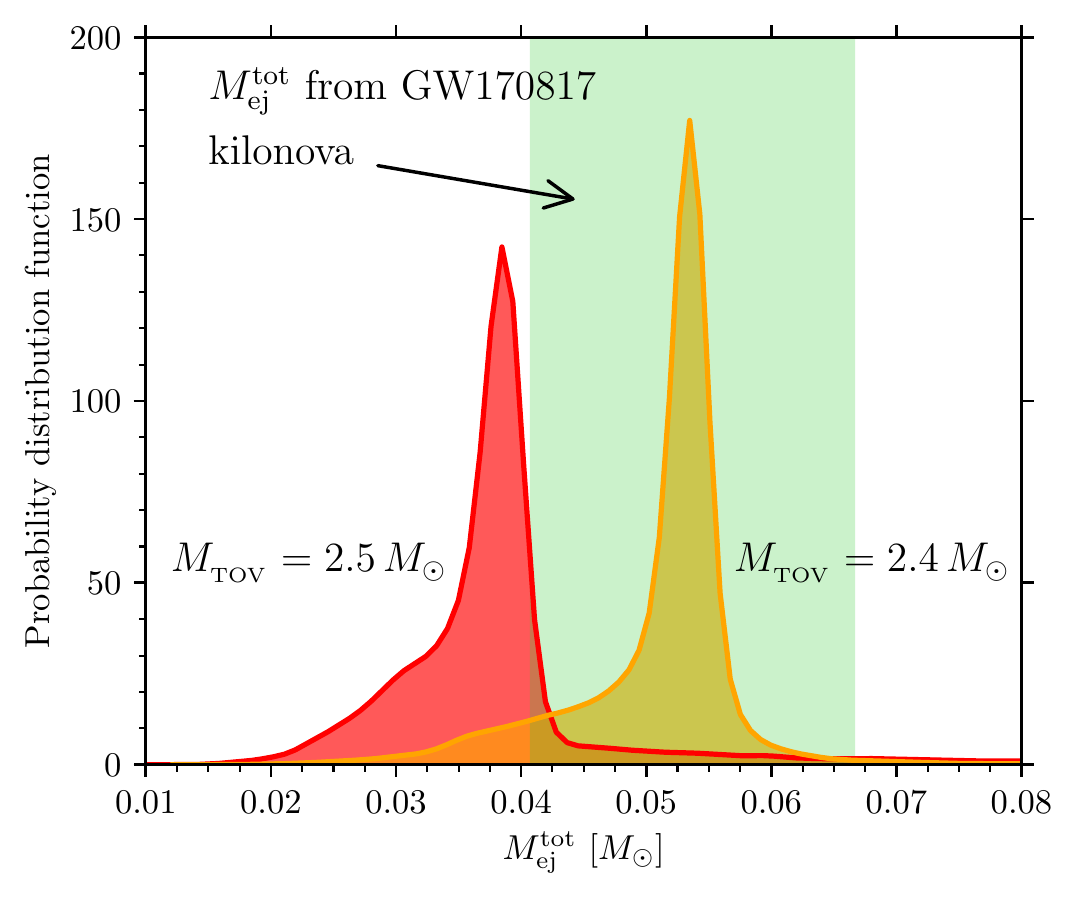}
  \caption{The same as Fig. \ref{fig:breakdown} but when we allow for the
    dimensionless spin to attain even smaller values, \ie $1.05 \leq \chi
    \lesssim 1.22$. Note that in this case disagreement in the radiated
    and ejected mass becomes even stronger for $M_{_{\rm
        TOV}}=2.5\,M_{\odot}$.}
  \label{fig:breakdown_2}
\end{figure*}

Finally, in Fig. \ref{fig:breakdown_2} we provide information that is
similar in content to that in Fig. \ref{fig:breakdown}, but when we allow
for the dimensionless spin to attain even smaller values, \ie $1.05 \leq
\chi \lesssim 1.22$. Note that in this case, the ejected mass for
$M_{_{\rm TOV}}=2.4\,M_{\odot}$ is within the observational bounds, but
the excess in radiated mass is more severe. The disagreement becomes even
stronger for $M_{_{\rm TOV}}=2.5\,M_{\odot}$.

As a concluding remark we note that the interpretation of the
  nature of GW190425 is likely unaffected by our findings on the maximum
  masses of neutron stars. While a BH-NS nature cannot be fully ruled
  out, the most plausible case of a NS-NS nature of the system is
  perfectly compatible with our findings on the maximum mass, as the
  initial masses in GW190425 are both well below the maximum-mass limit
  we have presented here \citep[see also][for a discussion on
    GW190425]{Most2020e}. On the other hand, an indirect impact that our
  results have on GW190425 is on whether the merger led to a prompt
  collapse (\ie where the hypermassive neutron star collapses to a black
  hole either at or shortly after merger), or to a stable long lived
  remnant. Using the results of \citet{Koeppel2019} \citep[but see
    also][]{Bauswein2017b}, and given the values for the maximum mass
  found here, a prompt or delayed collapse scenario seems likely for
  GW190425.

\end{document}